\documentclass[conference]{IEEEtran}
\IEEEoverridecommandlockouts
\makeatletter
\newcommand{\linebreakand}{%
\end{@IEEEauthorhalign}
\hfill\mbox{}\par
\mbox{}\hfill\begin{@IEEEauthorhalign}
}
\makeatother
\usepackage{cite}
\usepackage{amsmath,amssymb,amsfonts}
\usepackage{algorithmic}
\usepackage{graphicx}
\usepackage{textcomp}
\usepackage{xcolor}
\usepackage{booktabs}
\usepackage{multirow}
\usepackage{array}
\usepackage{url}
\usepackage{amsthm}
\usepackage{algorithm}

\newcommand{\method}{SAHC-NS}
\newcommand{\users}{\mathcal{U}}
\newcommand{\items}{\mathcal{I}}
\newcommand{\interactions}{\mathcal{E}}
\newcommand{\cands}{\mathcal{C}}

\newcommand{\emb}{\mathbf{e}}

\newcommand{\argmax}{\operatorname*{arg\,max}}

\begin{document}

\title{SAHC-NS: Structure-Aware and Hardness-Calibrated Negative Sampling for Implicit Collaborative Filtering}

\author{
	\IEEEauthorblockN{
		Jiayi Wu,
		Zhengyu Wu,
		Xunkai Li,
		Hongchao Qin,
		Rong-Hua Li,
		and Guoren Wang
	}
	\IEEEauthorblockA{
		Beijing Institute of Technology, Beijing 100811, China\\
		Corresponding author: Rong-Hua Li (lironghuabit@126.com)
	}
}

\maketitle

\begin{abstract}
	Negative sampling is a key component of implicit collaborative filtering (CF), as it enables recommenders to effectively learn user preferences.
	Existing negative sampling methods mostly follow a two-stage paradigm: they first construct a candidate negative pool for each user and then select negative samples from the pool according to predefined sampling rules.
	However, these methods usually overlook the hardness variation of candidate negative pools across users, making it difficult to adaptively adjust the hardness and informativeness of negative samples according to candidate-pool conditions.
	In addition, most existing samplers evaluate candidate negatives mainly through a matching score computed from the final aggregated user and item embeddings, while ignoring the structural differences captured by multi-hop neighborhood aggregation.
	As a result, the training value of negatives may be insufficiently characterized.
	To address these issues, we propose SAHC-NS, a Structure-Aware and Hardness-Calibrated Negative Sampling method.
	Specifically, SAHC-NS characterizes candidate negatives using their mean layer-wise matching strength and cross-layer structural discrepancy.
	This enables SAHC-NS to select informative negatives by taking cross-layer structural discrepancy into account, rather than relying solely on final matching scores.
	Moreover, SAHC-NS introduces a candidate-pool-aware hardness calibration module to dynamically adjust negative augmentation strength according to candidate-pool hardness, producing hardness-controllable negatives.
	Extensive experiments demonstrate the superiority of SAHC-NS over existing negative sampling methods.
\end{abstract}

\maketitle

\section{Introduction}
Implicit feedback~\cite{hu2008collaborative} is widely used in recommender systems because user behaviors such as clicks, purchases, and views are much easier to collect than explicit ratings~\cite{liang2021fedrec}.
However, implicit feedback only provides observed positive interactions, leaving a massive number of user-item pairs unobserved.
Since the number of unobserved interactions is usually extremely large, it is computationally impractical to use all of them for model training~\cite{liu2016relaxed}.
To address this issue, negative sampling has been widely adopted as an efficient training strategy for implicit collaborative filtering~\cite{ma2026negative}.
It constructs training instances by pairing each observed positive item with one or more sampled unobserved items, forming triplets such as $(u, i^+, j^-)$, where $i^+$ denotes a positive item and $j^-$ denotes a sampled negative item for user $u$.
With ranking or contrastive learning objectives such as Bayesian Personalized Ranking (BPR)~\cite{bpr} loss and Information Noise-Contrastive Estimation (InfoNCE)~\cite{oord2018representation} loss, the model is trained to pull user $u$ closer to the positive $i^+$ and push it away from the sampled negative $j^-$ in the embedding space.

Given the massive size of the unobserved item set, directly applying model-aware sampling criteria over all unobserved items is often computationally prohibitive, especially when such criteria involve computing user-item matching scores or performing embedding-level operations.
To improve sampling efficiency, many advanced negative sampling methods adopt a two-pass sampling paradigm~\cite{SRNS,GDNS,ANS}.
They first construct a relatively small candidate negative pool for each user by sampling or retrieving items from the unobserved set, and then select or synthesize more informative negatives from the pool according to specific sampling criteria.
For example, hard negative sampling methods~\cite{DNS, MixGCF,DNSMN} usually select negatives with high predicted preference scores from the candidate pool, since such negatives are more difficult to distinguish from positive items and thus provide stronger supervision signals.
Moreover, false-negative-aware methods~\cite{RealHNS, BNS, RecNS, wang2025exploration} usually incorporate false-negative risk estimates into the sample scores, reducing the chance of selecting potential positives as negatives and improving the accuracy of user preference learning.

Although the above two-pass sampling methods improve the informativeness or reliability of sampled negatives, they still suffer from two limitations.
\textbf{1. Candidate-pool Hardness Mismatch.}
Candidate negative pools can have substantially different hardness across training instances.
To verify this, we conduct a preliminary analysis on LightGCN~\cite{lightgcn} trained with DNS~\cite{DNS}, a representative two-pass negative sampling strategy.
For each positive interaction, we compute a normalized gap between the user-positive inner product and the highest user-negative inner product within the candidate pool.
As shown in Fig.~1 (a), candidate pools exhibit substantial hardness variation.
Specifically, 33.18\% and 31.69\% of candidate pools on Toys and Yelp have a normalized gap smaller than 0.10, indicating that they already contain negatives close to the corresponding positives.
By contrast, 44.47\% and 46.72\% of candidate pools have a normalized gap larger than 0.20, suggesting that many candidate pools are still dominated by relatively easy negatives.
However, existing augmentation-based~\cite{MixGCF, DivNS, RecNS} methods usually apply the same augmentation distribution to candidate pools with different hardness levels.
Such hardness-agnostic augmentation may make negatives from already hard pools overly difficult, thereby increasing potential false-negative risk, while failing to sufficiently enhance negatives from relatively easy pools and leaving them less informative for training.
\textbf{2.  Layer-wise Structural Blindness.}
Existing two-pass samplers usually select negatives from the candidate pool based on a matching score computed from the final aggregated embeddings.
However, this final-embedding-based score compresses the matching signals from different propagation layers into a single value.
Since different layers capture collaborative signals from different structural ranges, candidates with similar final scores may still exhibit distinct layer-wise matching behaviors.  
To empirically examine this issue, we further analyze whether candidate negatives with similar final-embedding-based scores have similar cross-layer structural discrepancies under the same preliminary setting.
Two negatives are considered score-similar if their final score difference is smaller than 0.01.
We then measure the absolute difference between their scale-normalized cross-layer discrepancies, where the discrepancy is computed as the standard deviation of layer-wise matching scores normalized by their average absolute magnitude.
As shown in Fig.~1 (b), many score-similar negative pairs still exhibit substantial normalized cross-layer discrepancy gaps.
Specifically, over 60\% of such pairs on both Toys and Yelp have discrepancy gaps larger than 0.20, and about one quarter exceed 0.50.
This suggests that the final-embedding-based score can mask structural differences captured by multi-hop neighborhood aggregation, leading to insufficient characterization of the training value of candidate negatives.

\begin{figure}
	\centering
	\includegraphics[width=\linewidth]{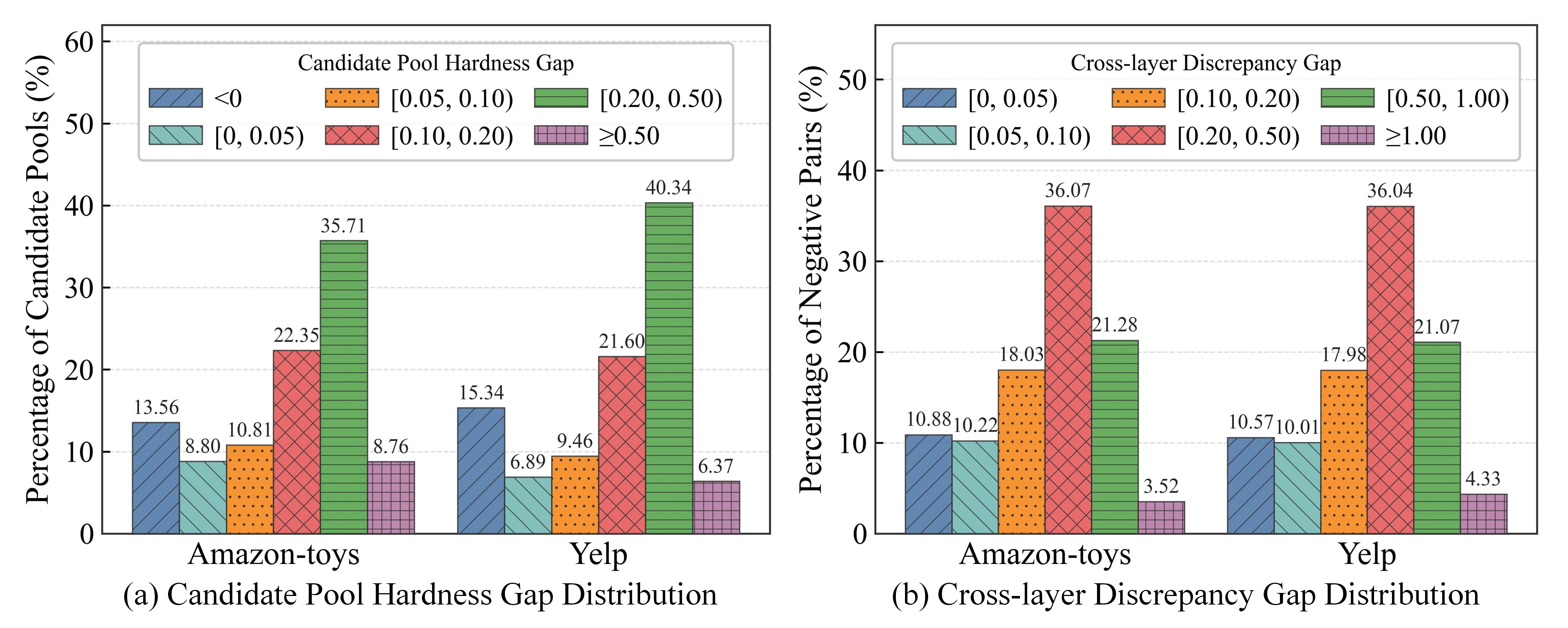}
	
	\vspace{-0.2cm}
	
	\caption{Motivation analysis of candidate-pool hardness and cross-layer structural discrepancy.}
	
	\vspace{-0.4cm}
	
\end{figure}

To address these problems, we propose \textbf{SAHC-NS}, a \textbf{S}tructure-\textbf{A}ware and \textbf{H}ardness-\textbf{C}alibrated \textbf{N}egative \textbf{S}ampling method.
SAHC-NS consists of two key components: structure-aware negative selection (SA) and candidate-pool-aware hardness calibration (HC).
SA uses the mean layer-wise matching score to characterize overall matching strength and scale-normalized cross-layer score variation to characterize structural discrepancy.
This allows SA to distinguish candidate negatives with similar final scores but different structural discrepancies, addressing the second limitation.
HC then adaptively calibrates the hardness augmentation strength according to the hardness of each candidate pool, addressing the first limitation and generating negatives with more controllable hardness.

The main contributions are summarized as follows:
\begin{itemize}[leftmargin=*]
	
	\item We design a structure-aware negative selection module that jointly considers mean hardness and cross-layer structural discrepancy derived from layer-wise matching scores, thereby selecting structurally informative negatives that may be overlooked by final-score-based hard negative sampling.
	
	\item We develop a candidate-pool-aware hardness calibration module to adapt negative augmentation to the hardness of different candidate pools. By dynamically adjusting the augmentation strength, this module avoids over-hardening negatives from already hard pools while enhancing negatives from relatively easy pools, thereby better balancing informativeness and reliability.
	
	\item We integrate the above two modules into SAHC-NS, a structure-aware and hardness-calibrated negative sampling method. Extensive experiments demonstrate the superiority of SAHC-NS over existing negative sampling baselines, and further ablation and analytical studies verify the effectiveness of its key components.
\end{itemize}

\section{Related Work}

\subsection{Implicit Collaborative Filtering}

Implicit collaborative filtering aims to learn user preferences from observed implicit feedback and recommend items by modeling historical user-item interaction patterns~\cite{wu2022survey}.
Representative early models include BPR-MF~\cite{bpr}, which optimizes pairwise rankings between positive and sampled negative items, and NeuMF~\cite{NCF}, which uses neural networks to model nonlinear user-item interactions.
Although these methods are effective, they rely on direct user-item interaction signals and have limited ability to explicitly exploit the high-order collaborative structure of the user-item interaction graph.
Motivated by this, GNN-based CF methods have become widely used, as they can propagate user and item representations over the interaction graph to capture high-order collaborative signals~\cite{gao2023survey}.
NGCF~\cite{NGCF} incorporates high-order connectivity through message propagation on the user-item bipartite graph, while LightGCN~\cite{lightgcn} simplifies graph convolution by retaining only neighborhood aggregation.
Further studies extend GNN-based recommendation in different directions, such as LR-GCCF~\cite{LR-GCCF} with residual propagation, DGCF~\cite{DGCF} with intent disentanglement, and SGL~\cite{SGL}, SimGCL~\cite{SimGCL}, and XSimGCL~\cite{XSimGCL} with self-supervised contrastive learning.
Despite their different model architectures, these models are typically trained with random negative sampling, where unobserved items are uniformly sampled as negatives.
This simple strategy makes model training efficient, but the quality of sampled negatives can significantly affect the learned user-item embeddings, motivating extensive studies on negative sampling in implicit CF.

\subsection{Negative Sampling in Implicit Collaborative Filtering}

Negative sampling is widely used in implicit CF to select informative negatives from unobserved items~\cite{yang2024does}.
Early methods adopt static strategies, such as random sampling~\cite{bpr} and popularity-based sampling~\cite{rendle2014improving, mikolov2013distributed}, which draw negatives from predefined distributions without considering the current model state.
Although simple and efficient, these methods often sample easy negatives and thus provide limited training signals.
To improve sampling efficiency and sample informativeness, many advanced methods adopt a two-pass sampling paradigm.
They first construct a small candidate negative pool, and then select or synthesize more informative negatives from this pool according to model-aware sampling criteria.
Within this paradigm, hard negative sampling methods, such as DNS~\cite{DNS} and DNS(M,N)~\cite{DNSMN}, select negative items with high predicted preference scores from the candidate pool, thereby strengthening the model's ability to distinguish positives from challenging negatives.
Beyond selecting real negative items, augmentation-based methods synthesize harder negative embeddings in the embedding space.
For example, MixGCF~\cite{MixGCF} generates hard negative embeddings for graph neural network-based recommenders through embedding-level mixing, while ANS~\cite{ANS} synthesizes hard negatives by disentangling negative embeddings and augmenting their easy factors.
DENS~\cite{DENS} improves negative sampling by disentangling relevant and irrelevant factors, and selecting negatives that are close to positives in irrelevant factors but distinct in relevant factors.
Considering that overly hard negatives may have a higher risk of being false negatives, another line of work focuses on improving the reliability of negative sampling.
SRNS~\cite{SRNS} and GDNS~\cite{GDNS} exploit training dynamics, such as the variance of sample losses, to reduce the chance of selecting potential false negatives.
RealHNS~\cite{RealHNS} and RecNS~\cite{RecNS} leverage auxiliary signals such as cross-domain information or exposed-but-unclicked items to identify reliable negatives.
AHNS~\cite{AHNS} controls negative hardness according to positive-sample information, thereby alleviating the false-negative risk.
In addition, DivNS~\cite{DivNS} maintains hard-negative caches and applies diversity-augmented \(k\)-DPP~\cite{kulesza2011kdpp} sampling to select diverse negatives.

Despite their effectiveness, the above two-pass sampling methods usually ignore the hardness variation among different candidate pools, which may cause over-hardening for negatives from already hard pools and insufficient enhancement for negatives from relatively easy pools.
The former may increase the potential false-negative risk, while the latter may lead to less informative training signals.
In contrast, SAHC-NS adaptively calibrates negative augmentation according to candidate-pool hardness, thereby preventing generated negatives from being either overly hard or insufficiently informative.
Additionally, existing methods mainly rely on matching scores computed from final aggregated embeddings to evaluate candidate negatives.
However, such final-score-based characterization may neglect the cross-layer structural discrepancy captured by multi-hop neighborhood aggregation, leading to an incomplete estimation of negative training value.
In contrast, SAHC-NS characterizes candidate negatives using mean hardness and scale-normalized cross-layer structural discrepancy derived from their layer-wise matching scores, enabling multi-hop structure-aware negative selection.

\begin{figure*}
	\centering
	\includegraphics[width=\linewidth]{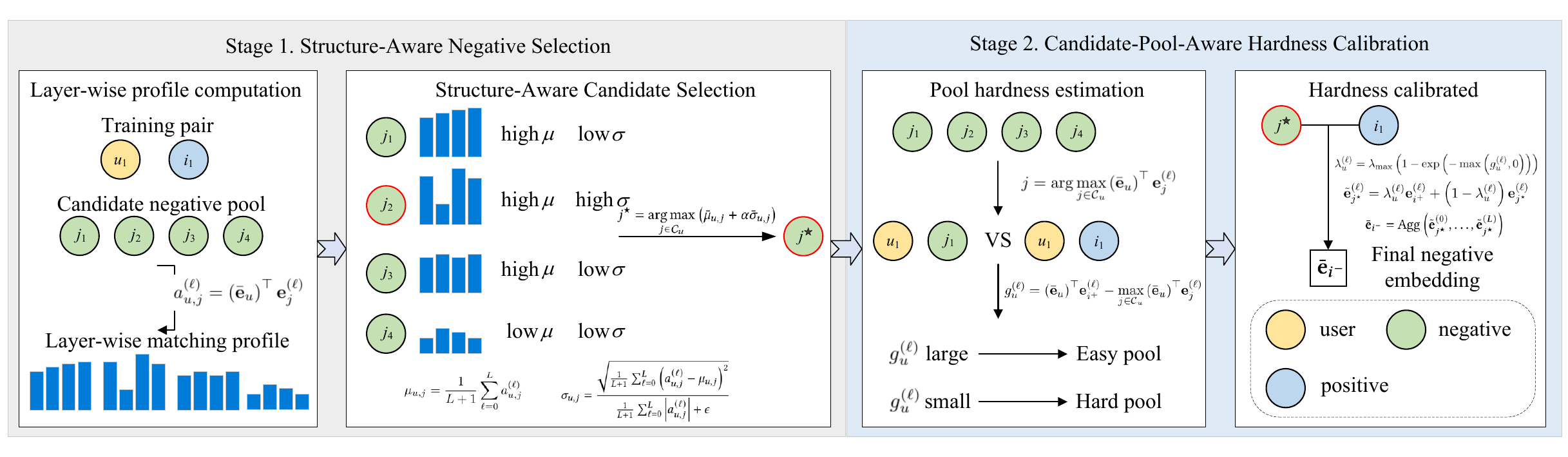}
	
	\vspace{-0.1cm}
	
	\caption{Illustration of SAHC-NS for a training pair and its candidate negative pool.}
	
	\vspace{-0.2cm}
	
\end{figure*}

\section{Methodology}
\label{sec:method}

This section first formulates the negative sampling problem, and then introduces the proposed SAHC-NS method. As shown in Fig.~2, \method~consists of two modules: structure-aware negative selection (SA) and candidate-pool-aware hardness calibration (HC).
The SA module first computes the layer-wise matching profile for each candidate within the candidate pool, i.e., the sequence of matching scores across propagation layers, and then performs structure-aware negative selection by jointly considering its overall matching strength and cross-layer structural discrepancy.
The HC module first estimates candidate-pool hardness based on the user-positive pair and the candidate pool, and then derives layer-wise calibration strengths to adjust the negative selected by SA.

\subsection{Problem Formulation}

Let $\users$ and $\items$ denote the user and item sets. Implicit feedback data are represented by an interaction set $\interactions$, where $(u,i)\in\interactions$ indicates that user $u$ has interacted with item $i$. For each user $u$, let $I_u^+$ be the observed item set and $I_u^- = \items\setminus I_u^+$ be the unobserved item set. 

In negative-sampling-based pairwise training, each observed interaction $(u,i^+)$ is paired with a negative embedding generated by the negative sampling method:
\begin{equation}
	\bar{\emb}_{i^-} = \pi(u,i^+,\cands_{u},\mathcal{R})
\end{equation}
where $\cands_{u}=\{j_1,\ldots,j_N\}\subset I_u^-$ denotes the candidate negative pool drawn from the unobserved item set of user $u$, $\mathcal{R}$ denotes the user and item embeddings produced by the recommender, and $\bar{\emb}_{i^-}$ is the negative embedding used for training.
For samplers that return a real item, $\bar{\emb}_{i^-}$ is the embedding of the selected item.
For synthesis-based samplers, it can be an augmented representation in the embedding space.

Since SAHC-NS relies on layer-wise representations, it is designed for GNN-based implicit CF recommenders~\cite{lightgcn,NGCF,SimGCL}, which learn user and item embeddings through layer-wise neighborhood propagation.
Given an interaction graph, we denote the layer-wise embeddings of user $u$ and item $i$ as
\begin{equation}
	\{\emb_u^{(0)},\emb_u^{(1)},\ldots,\emb_u^{(L)}\}, \quad
	\{\emb_i^{(0)},\emb_i^{(1)},\ldots,\emb_i^{(L)}\},
\end{equation}
where different layers reflect collaborative signals from different ranges of neighborhood propagation.
The final embedding is obtained by a recommender-specific aggregation function:
\begin{equation}
	\bar{\emb}_{u}= \operatorname{Agg}(\emb_u^{(0)},\ldots,\emb_u^{(L)}),
	\quad
	\bar{\emb}_{i}= \operatorname{Agg}(\emb_i^{(0)},\ldots,\emb_i^{(L)}),
\end{equation}
and the final preference score is computed as $\hat{y}_{ui}=\bar{\emb}_u^\top \bar{\emb}_i.$
Given the final negative embedding $\bar{\emb}_{i^-}$ obtained by a negative sampling strategy, the BPR objective can be written as
\begin{equation}
	\mathcal{L}_{\mathrm{BPR}}
	=
	- \sum_{(u,i^+)}
	\log \phi(\hat{y}_{ui^+}-\hat{y}_{ui^-})
	+
	\beta \lVert \Theta \rVert_2^2,
	\label{eq:bpr}
\end{equation}
where $\phi(\cdot)$ denotes the sigmoid function, 
$\hat{y}_{ui^-} \! = \! \bar{\emb}_{u}^{\top}\bar{\emb}_{i^-}$ is the negative preference score, 
$\Theta$ denotes model parameters, and $\beta$ is a regularization coefficient.
\method~does not modify the model architecture or the loss; it modifies the sampling strategy $\pi(\cdot)$ by deciding which candidate should be selected and how the augmentation strength of the negative embedding should be calibrated.

This formulation leads to two design requirements.
First, since GNN-based recommenders produce layer-wise embeddings that reflect collaborative signals from different ranges of neighborhood propagation, candidate negatives should not be evaluated solely by matching scores computed from final aggregated embeddings.
Instead, their layer-wise matching behavior should be considered to capture cross-layer structural discrepancy.
Second, since candidate pools may exhibit different hardness levels, sampling augmentation strengths from a globally fixed distribution may over-harden negatives from already hard pools or leave negatives from relatively easy pools insufficiently informative. Therefore, the augmentation strength should be adaptively calibrated according to candidate-pool hardness.
Guided by these two requirements, we introduce a structure-aware and hardness-calibrated negative sampling method in the following subsections.

\subsection{Structure-Aware Negative Selection}
Existing two-pass samplers~\cite{DNS, DivNS, DNSMN} usually select negatives from the candidate pool based on matching scores computed from final aggregated embeddings, which compress the matching signals from different propagation layers into a single representation.
However, candidates with similar final matching scores may still exhibit different layer-wise matching behaviors, reflecting different multi-hop collaborative structures.
To capture such cross-layer structural discrepancy, SA evaluates candidates through their layer-wise matching profiles and then selects structurally informative negatives.
Considering that the aggregated user embedding provides a stable preference anchor, we use it as the user-side scoring representation and compare it with the item representation at each propagation layer.
Specifically, let $\bar{\emb}_{u}$ denote the aggregated user representation.
For each candidate negative $j\in\cands_u$, SA computes its layer-wise matching score as:
\begin{equation}
	a_{u,j}^{(\ell)} =
	\left(\bar{\emb}_u\right)^\top \emb_j^{(\ell)},
	\quad \ell=0,\ldots,L.
	\label{eq:layer_score}
\end{equation}
The resulting vector
$\mathbf{a}_{u,j}
=
[a_{u,j}^{(0)},a_{u,j}^{(1)},\ldots,a_{u,j}^{(L)}]$
is the layer-wise matching profile of candidate $j$.
Since item embeddings at different propagation layers encode collaborative signals from different structural ranges, this profile characterizes how the candidate matches the user's aggregated preference across different item-side structural views.
Thus, it preserves layer-wise matching variations that may be hidden by the final aggregated score, providing the basis for subsequently measuring cross-layer structural discrepancy.

Based on this profile, the SA module summarizes each candidate with two statistics: mean hardness and cross-layer structural discrepancy.
Mean hardness is defined as
\begin{equation}
	\mu_{u,j}
	=
	\frac{1}{L+1}\sum_{\ell=0}^{L} a_{u,j}^{(\ell)},
	\label{eq:mu}
\end{equation}
which measures the candidate's average matching strength across propagation layers.
It preserves the main intuition of hard negative sampling by favoring candidates that are closer to the user's aggregated preference representation under multiple item-side collaborative views, thereby providing stronger supervision for distinguishing positives from challenging negatives. 
Notably, for mean-aggregation backbones such as LightGCN, $\mu_{u,j}$ is algebraically equivalent to the conventional matching score computed from the final aggregated item embedding. 
In such cases, the mean term preserves the conventional hardness signal, while the cross-layer structural discrepancy provides complementary information beyond final-score-based negative selection.

Cross-layer structural discrepancy is defined as
\begin{equation}
	\sigma_{u,j}
	=
	\frac{
		\sqrt{
			\frac{1}{L+1}\sum_{\ell=0}^{L}
			\left(a_{u,j}^{(\ell)}-\mu_{u,j}\right)^2
		}
	}{
		\frac{1}{L+1}\sum_{\ell=0}^{L}
		\left|a_{u,j}^{(\ell)}\right|+\epsilon
	}.
	\label{eq:sigma}
\end{equation}
Here, $\epsilon$ is a small constant for numerical stability.
It measures the relative variation of the candidate's matching behavior across propagation layers. 
The numerator captures the cross-layer fluctuation of matching scores, while the denominator normalizes this fluctuation by the candidate's average absolute matching magnitude. 
This normalization reduces the influence of candidate-specific score scales, allowing the discrepancy to better reflect relative changes in matching behavior across different propagation depths. 
A large $\sigma_{u,j}$ indicates that the candidate exhibits pronounced inconsistency in matching behavior across propagation layers.
For example, a candidate may match the user's aggregated preference strongly through its shallow-layer item representation but become less similar through its higher-order item representation, or vice versa.
Since different layers reflect collaborative signals from different structural ranges, such fluctuation suggests that the candidate lies in a structurally ambiguous region: it may be difficult to distinguish from positives under some collaborative views, but not under others.
These candidates are informative because they expose preference boundaries that are hidden by the final aggregated score.
Therefore, $\sigma_{u,j}$ complements mean hardness by capturing cross-layer structural discrepancy that is hidden in average matching strength, enabling the sampler to identify structurally informative negatives.

Considering that mean hardness and structural discrepancy may have different numerical scales, directly combining $\mu_{u,j}$ and $\sigma_{u,j}$ may cause one statistic to dominate the selection score. We therefore further standardize both statistics within each candidate pool:
\begin{equation}
	\begin{aligned}
		\bar{\mu}_{u,j}
		& \! = \!
		\frac{\mu_{u,j}\!-\!\operatorname{mean}_{k\in\cands_u}\!\mu_{u,k}}
		{\operatorname{std}_{k\in\cands_u}\!\mu_{u,k}+\epsilon}, \! \!
		&
		\bar{\sigma}_{u,j}
		& \! = \!
		\frac{\sigma_{u,j}\!-\!\operatorname{mean}_{k\in\cands_u}\!\sigma_{u,k}}
		{\operatorname{std}_{k\in\cands_u}\!\sigma_{u,k}+\epsilon}.
	\end{aligned}
	\label{eq:norm_mu_sigma}
\end{equation}
This pool-level standardization makes the two statistics comparable for relative candidate ranking within the current candidate pool.
The final selection score is
\begin{equation}
	q_{u,j}
	=
	\bar{\mu}_{u,j} + \alpha \bar{\sigma}_{u,j},
	\label{eq:q}
\end{equation}
where $\alpha\geq 0$ controls the contribution of cross-layer structural discrepancy. The selected candidate is
\begin{equation}
	j^\star = \argmax_{j\in\cands_u} q_{u,j}.
	\label{eq:select}
\end{equation}
Through this scoring function, the SA module selects negatives that are not only hard on average but also exhibit informative structural discrepancies across propagation layers.

\subsection{Candidate-Pool-Aware Hardness Calibration}

The SA module selects an informative negative by considering both mean hardness and cross-layer structural discrepancy.
However, the selected candidate is still constrained by the candidate pool.
If the candidate pool is dominated by easy negatives, even the best candidate selected by SA may still provide limited training signals.
Conversely, if the pool already contains negatives close to the positive item, applying strong augmentation may over-harden the generated negative and increase potential false-negative risk.
Therefore, HC first estimates the hardness of the candidate pool and then calibrates the augmentation strength accordingly.
Specifically, under the same aggregated user representation, it compares the positive with the highest-scoring candidate negative in the pool at each propagation layer.
If the highest-scoring candidate is already close to the positive item, the pool can be regarded as sufficiently hard; otherwise, a large positive-negative gap indicates that the pool is still dominated by relatively easy negatives.
This provides a direct estimate of the hardness level of the candidate pool.

Let the positive score at layer $\ell$ be
$p_{u,i^+}^{(\ell)}
=
\left(\bar{\emb}_u\right)^\top \emb_{i^+}^{(\ell)}$
and let the strongest candidate score in the pool be
$
m_u^{(\ell)}
=
\max_{j\in\cands_u} a_{u,j}^{(\ell)}.
$
We use the maximum score over the entire candidate pool rather than the score of the selected candidate $j^\star$, because $j^\star$ reflects both hardness and cross-layer structural discrepancy, whereas HC aims to estimate the pool hardness itself. The highest-scoring candidate therefore provides a more direct indicator of how hard the candidate pool already is.
The positive-negative gap is then defined as
\begin{equation}
	g_u^{(\ell)}
	=
	p_{u,i^+}^{(\ell)} - m_u^{(\ell)}.
	\label{eq:gap}
\end{equation}
A larger gap indicates that even the highest-scoring candidate in the pool is still far from the positive item, suggesting an easier candidate pool.
By contrast, when $g_u^{(\ell)}$ is close to or smaller than zero, the pool already contains a sufficiently hard candidate whose score is comparable to or even higher than that of the positive item, indicating that the current pool is already hard.
We map this gap to a bounded pool-hardness score:
\begin{equation}
	\rho_u^{(\ell)}
	=
	\exp\left(-\max\left(g_u^{(\ell)},0\right)\right).
	\label{eq:quality}
\end{equation}
When the pool is hard, $g_u^{(\ell)}\leq 0$ and thus $\rho_u^{(\ell)}=1$.
When the pool is easy, $\rho_u^{(\ell)}$ decays smoothly as the positive-negative gap increases.
The exponential form avoids division by the positive score, which can be unstable early in training, and maps the BPR-style positive-negative margin to a smooth pool-hardness score.

The layer-wise calibration strength is then defined as
\begin{equation}
	\lambda_u^{(\ell)}
	=
	\lambda_{\max}\left(1-\rho_u^{(\ell)}\right),
	\quad 0\leq \lambda_{\max}\leq 1.
	\label{eq:lambda}
\end{equation}
where $\lambda_{\max}$ controls the maximum calibration strength, preventing the calibrated negative from being moved excessively toward the positive item and thus mitigating potential false-negative risk.
The coefficient is computed layer by layer according to the estimated hardness of the current candidate pool, since its hardness can vary across propagation layers.
Layer-wise calibration therefore strengthens layers where the pool is relatively easy, while preserving layers where the original candidates are already hard.

For the selected candidate $j^\star$, HC constructs the calibrated negative representation at each layer:
\begin{equation}
	\tilde{\emb}_{j^\star}^{(\ell)}
	=
	\lambda_u^{(\ell)} \emb_{i^+}^{(\ell)}
	+
	\left(1-\lambda_u^{(\ell)}\right)
	\emb_{j^\star}^{(\ell)}.
	\label{eq:mix}
\end{equation}
Notably, while the interpolation form is related to MixGCF~\cite{MixGCF}, MixGCF samples the mixing strength from a predefined distribution, whereas HC adaptively determines $\lambda_u^{(\ell)}$ according to the estimated hardness of the current candidate pool. This enables HC to calibrate hardness enhancement to different candidate-pool conditions.
During optimization, $\lambda_u^{(\ell)}$ is treated as a non-differentiable calibration quantity, while gradients are retained for the positive embedding $\emb_{i^+}^{(\ell)}$ and selected negative embedding $\emb_{j^\star}^{(\ell)}$ in Eq.~\eqref{eq:mix}.

Following the aggregation rule of the implicit CF recommender, SAHC-NS aggregates
$\{\tilde{\emb}_{j^\star}^{(\ell)}\}_{\ell=0}^{L}$
into the final negative embedding
\begin{equation}
	\bar{\emb}_{i^-}
	=
	\operatorname{Agg}
	\left(
	\tilde{\emb}_{j^\star}^{(0)},
	\ldots,
	\tilde{\emb}_{j^\star}^{(L)}
	\right)
	\label{eq:negative_agg}
\end{equation}
which is then used to compute the negative preference score
$\hat{y}_{ui^-}=\bar{\emb}_{u}^{\top}\bar{\emb}_{i^-}$ in the BPR objective.

Algorithm~\ref{alg:sahc} describes how SAHC-NS generates negative embeddings for mini-batch pairs. (Lines 1-3) obtain the layer-wise embeddings required by the current mini-batch. (Lines 5-9) perform structure-aware negative selection by considering both mean hardness and cross-layer structural discrepancy (Lines 10-13) conduct candidate-pool-aware hardness calibration to adaptively construct the negative embedding. The resulting negative embeddings are then returned for subsequent pairwise optimization.

\subsection{Time Complexity}

The time complexity of \method~mainly arises from two modules: structure-aware negative selection (SA) and candidate-pool-aware hardness calibration (HC).
For the SA module, let $B$ be the batch size, $N$ be the candidate-pool size, $L$ be the number of propagation layers, and $d$ be the embedding dimension. Computing layer-wise matching scores for all candidates costs $\mathcal{O}(BN(L+1)d)$. The subsequent computation of mean hardness, cross-layer structural discrepancy, normalization statistics, and selection scores only involves scalar operations over layer-wise scores, with a complexity of $\mathcal{O}(BN(L+1))$. Thus, the complexity of SA is $\mathcal{O}(BN(L+1)d)$.
For the HC module, computing the positive-negative gaps, pool-hardness scores, and layer-wise calibration strengths costs $\mathcal{O}(BN(L+1))$. After the candidate $j^\star$ is selected, constructing its calibrated negative representation costs $\mathcal{O}(B(L+1)d)$. Therefore, the overall time complexity of \method~can be approximated as $\mathcal{O}(BN(L+1)d)$.

\begin{algorithm}[t]
	\renewcommand{\algorithmicrequire}{\textbf{Input:}}
	\renewcommand{\algorithmicensure}{\textbf{Output:}}
	\begin{algorithmic}[1]
		\REQUIRE interaction graph $\mathcal{G}$, Mini-batch pairs $\mathcal{B}=\{(u,i^+)\}$, candidate pools $\{\cands_u\}$, recommender $f_\Theta$, structural discrepancy weight $\alpha$, maximum calibration strength $\lambda_{\max}$
		\ENSURE Negative embeddings 
		$\widetilde{\mathcal{E}}_{\mathcal{B}}^{-}$ generated by SAHC-NS for subsequent pairwise optimization.
		
		\STATE Initialize $\widetilde{\mathcal{E}}_{\mathcal{B}}^{-}\leftarrow\emptyset$.
		\STATE Perform a forward propagation of $f_\Theta$ on $\mathcal{G}$ to obtain layer-wise user and item embeddings.
		\STATE Index $\bar{\emb}_u$, $\emb_{i^+}^{(0:L)}$, and $\{\emb_j^{(0:L)}\}_{j\in\cands_u}$ for each $(u,i^+)\!\in\!\mathcal{B}$.
		
		\FOR{each training pair $(u,i^+)\in\mathcal{B}$}
		\STATE \textbf{Structure-aware negative selection.}
		\STATE \hspace{\algorithmicindent}Build the layer-wise matching profiles between $u$ and candidates in $\cands_u$ based on $\bar{\emb}_u$ and $\!\{\emb_j^{(0:L)}\}\!_{j\in\cands_u}$ by~\eqref{eq:layer_score}.

		\STATE \hspace{\algorithmicindent}Derive candidate-level mean hardness $\mu_{u,j}$ and scale-normalized cross-layer structural discrepancy $\sigma_{u,j}$ from the layer-wise matching profiles by Eqs.~\eqref{eq:mu} and \eqref{eq:sigma}.
		\STATE \hspace{\algorithmicindent}Standardize $\mu_{u,j}$ and $\sigma_{u,j}$ within $\cands_u$ to obtain $\bar{\mu}_{u,j}$ and $\bar{\sigma}_{u,j}$ by Eq.~\eqref{eq:norm_mu_sigma}.
		\STATE \hspace{\algorithmicindent}Score candidates using $\bar{\mu}_{u,j}$ and $\bar{\sigma}_{u,j}$, and select $j^\star$ by Eqs.~\eqref{eq:q} and \eqref{eq:select}.
		
		\STATE \textbf{Candidate-pool-aware hardness calibration.}
		
		\STATE \hspace{\algorithmicindent}Compute the layer-wise positive-negative gap $g_u^{(\ell)}$ based on $\bar{\emb}_u$, $\emb_{i^+}^{(0:L)}$, and $\{\emb_j^{(0:L)}\}_{j\in\cands_u}$ by Eq.~\eqref{eq:gap}.
		
		\STATE \hspace{\algorithmicindent}Compute the layer-wise calibration strength $\lambda_u^{(\ell)}$ based on $g_u^{(\ell)}$ and $\lambda_{\max}$ by Eqs.~\eqref{eq:quality} and \eqref{eq:lambda}.
		\STATE \hspace{\algorithmicindent}Obtain the negative embedding $\tilde{\emb}_{j^\star}^{(0:L)}$ based on $\lambda_u^{(\ell)}$, $\emb_{i^+}^{(0:L)}$, and $\emb_{j^\star}^{(0:L)}$ by Eq.~\eqref{eq:mix}.
		\STATE \hspace{\algorithmicindent}Aggregate $\{\tilde{\emb}_{j^\star}^{(\ell)}\}_{\ell=0}^{L}$ into the final negative embedding $\bar{\emb}_{i^-}$ by Eq.~\eqref{eq:negative_agg}, and add $\bar{\emb}_{i^-}$ to $\widetilde{\mathcal{E}}_{\mathcal{B}}^{-}$.
		
		\ENDFOR
		\STATE \textbf{return} $\widetilde{\mathcal{E}}_{\mathcal{B}}^{-}$.
	\end{algorithmic}
	\caption{Structure-Aware Hardness-Calibrated Negative Sampling for Mini-Batch Pairwise Training.}
	\label{alg:sahc}
\end{algorithm}

\section{Experiments}
\label{sec:exp}

To demonstrate the effectiveness of our proposed SAHC-NS, we conducted extensive experiments and answered the following questions: 
\textbf{(RQ1)} How does SAHC-NS perform in comparison to existing negative sampling methods? \textbf{(RQ2)} How do the key hyperparameters of SAHC-NS affect recommendation performance? \textbf{(RQ3)} How do the SA and HC modules contribute to the effectiveness of SAHC-NS, and what mechanisms explain their improvements?
\textbf{(RQ4)} How well does SAHC-NS perform when integrated with various GNN-based implicit CF models?

\subsection{Experimental Setup}

\subsubsection{Datasets}

We selected three widely used real-world datasets (Amazon-toys, Yelp, and ML-1m) to evaluate the effectiveness of SAHC-NS. 
Amazon-toys\footnote{\url{https://cseweb.ucsd.edu/~jmcauley/datasets/amazon_v2/}} is collected from the Amazon review platform and contains user-item interactions in the Toys and Games category, spanning the period from 2014 to 2018.
Yelp\footnote{\url{https://business.yelp.com/data/resources/open-dataset/}} contains user interaction records from the Yelp platform spanning from 2015 to 2018. 
ML-1m\footnote{\url{https://grouplens.org/datasets/movielens/1m/}}  is derived from user ratings for movies on the Movielens website.
Following prior work~\cite{NGCF, lightgcn, MCNS}, we adopt the 10-core setting for all datasets and randomly split the interactions into training, validation, and test sets with a ratio of 8:1:1.
The dataset statistics are reported in Table~\ref{tab:datasets}.
We evaluate performance using Recall@k and NDCG@k with $k\in\{10,20\}$.

\begin{table}[t]
	\centering
	\caption{Dataset Statistics.}
	
	\vspace{-0.1cm}
	
	\label{tab:datasets}
	\begin{tabular}{ccccc}
		\toprule
		Dataset & Users & Items & Interactions & Density \\
		\midrule
		Amazon-toys & 15,306   & 10,074   & 247,181  & 0.160\%  \\
		Yelp & 20,206  & 15,674  & 468,033 & 0.148\% \\
		ML-1m &  6,040& 3,629&836,478&3.816\% \\
		\bottomrule
	\end{tabular}
	
	\vspace{-0.2cm}
	
\end{table}

\newcommand{\std}[1]{{\scriptsize$\pm$#1}}

\begin{table*}[t]  
	\centering  
	\caption{Performance comparison on three real-world datasets (\%). 
		The best and second-best results are highlighted in \textbf{bold} and \underline{underlined}. 
		Statistically significant improvements over the strongest baseline are marked with $^{*}$ for $p<0.05$.} 
	
	\vspace{-0.2cm}
	
	\setlength{\tabcolsep}{0.48mm} 
	\label{tab:main}  
	\begin{tabular}{c|cccc|cccc|cccc}  
		\toprule  
		\multirow{1}{*}{Dataset}  
		& \multicolumn{4}{c|}{Amazon-toys}  
		& \multicolumn{4}{c|}{Yelp}  
		& \multicolumn{4}{c}{ML-1m} \\  
		\cmidrule(r){1-1} \cmidrule(lr){2-5}\cmidrule(lr){6-9}\cmidrule(lr){10-13}  
		\multirow{1}{*}{Method}   
		& R@10 & N@10 & R@20 & N@20  
		& R@10 & N@10 & R@20 & N@20  
		& R@10 & N@10 & R@20 & N@20 \\  
		\midrule  
		
		RNS        & 7.88\std{0.17} & 5.11\std{0.08} & 11.32\std{0.19} & 6.13\std{0.07} 
		& 8.11\std{0.10} & 5.24\std{0.09} & 13.28\std{0.09} & 6.86\std{0.07} 
		& 17.30\std{0.24} & 20.77\std{0.22} & 26.64\std{0.12} & 22.85\std{0.14} \\ 
		
		DNS        & 9.09\std{0.13} & 6.08\std{0.08} & 12.66\std{0.20} & 7.14\std{0.09} 
		& 8.39\std{0.08} & 5.49\std{0.06} & 13.67\std{0.06} & 7.15\std{0.06} 
		& 18.19\std{0.15} & 21.60\std{0.16} & 27.68\std{0.24} & 23.75\std{0.18} \\ 
		
		DNS(M,N)   & 9.30\std{0.10} & \underline{6.38\std{0.10}} & 12.64\std{0.17} & \underline{7.39\std{0.14}} 
		& 8.73\std{0.10} & 5.69\std{0.10} & 13.90\std{0.13} & 7.32\std{0.11} 
		& 18.39\std{0.13} & 21.83\std{0.16} & 28.20\std{0.13} & 24.03\std{0.14} \\ 
		
		MixGCF     & \underline{9.44\std{0.15}} & 6.29\std{0.06} & \underline{13.02\std{0.21}} & 7.36\std{0.08} 
		& \underline{8.84\std{0.06}} & \underline{5.77\std{0.05}} & \underline{14.34\std{0.15}} & \underline{7.50\std{0.07}} 
		& \underline{18.67\std{0.11}} & \underline{22.33\std{0.07}} & \underline{28.38\std{0.11}} & \underline{24.12\std{0.05}} \\ 
		
		DENS   & 8.31\std{0.15} & 5.56\std{0.19} & 12.14\std{0.12} & 6.69\std{0.17} 
		& 8.56\std{0.05} & 5.59\std{0.10} & 13.81\std{0.12} & 7.25\std{0.08} 
		& 17.55\std{0.19} & 20.67\std{0.14} & 26.33\std{0.20} & 22.50\std{0.13} \\ 
		
		BNS        & 8.93\std{0.11} & 5.83\std{0.07} & 12.56\std{0.05} & 6.91\std{0.05} 
		& 8.52\std{0.03} & 5.50\std{0.05} & 13.67\std{0.12} & 7.13\std{0.07} 
		& 18.46\std{0.26} & 22.07\std{0.13} & 27.60\std{0.21} & 23.94\std{0.07} \\ 
		
		AHNS       & 8.95\std{0.11} & 5.94\std{0.07} & 12.46\std{0.12} & 6.99\std{0.07} 
		& 8.32\std{0.17} & 5.45\std{0.09} & 13.65\std{0.20} & 7.12\std{0.10} 
		& 18.06\std{0.31} & 21.54\std{0.19} & 27.33\std{0.30} & 23.57\std{0.20} \\ 
		
		DivNS      & 8.89\std{0.19} & 5.87\std{0.11} & 12.46\std{0.31} & 6.94\std{0.15} 
		& 8.43\std{0.13} & 5.50\std{0.07} & 13.69\std{0.14} & 7.15\std{0.08} 
		& 17.95\std{0.29} & 21.27\std{0.23} & 27.32\std{0.21} & 23.40\std{0.20} \\ 
		
		\method    & \textbf{9.61\std{0.14}}$^{*}$ & \textbf{6.57\std{0.08}}$^{*}$ & \textbf{13.24\std{0.22}}$^{*}$ & \textbf{7.65\std{0.12}}$^{*}$
		& \textbf{9.03\std{0.04}}$^{*}$ & \textbf{5.95\std{0.03}}$^{*}$ & \textbf{14.63\std{0.17}}$^{*}$ & \textbf{7.72\std{0.06}}$^{*}$ 
		& \textbf{18.96\std{0.17}}$^{*}$ & \textbf{22.50\std{0.15}}$^{*}$ & \textbf{28.63\std{0.15}}$^{*}$ & \textbf{24.56\std{0.18}}$^{*}$ \\ 
		
		\bottomrule  
	\end{tabular}  
	
	\vspace{-0.2cm}  
\end{table*}

\subsubsection{Baselines}

We compare SAHC-NS with eight representative and competitive negative sampling baselines, including RNS, DNS, DNS(M,N), MixGCF, DENS, BNS, AHNS, and DivNS. 
For a fair comparison, unless otherwise specified for a particular method, the candidate-pool size is fixed to 10 for two-pass negative sampling methods that explicitly construct a candidate pool, including SAHC-NS. 
For other method-specific hyperparameters, we adopt the configurations reported in the original papers or official implementations when the corresponding baselines were evaluated on the same datasets as ours. 
For datasets not covered by the original studies or official implementations, we tune these hyperparameters based on validation performance. 
The details of these baselines and their method-specific hyperparameters are summarized below:

\begin{itemize}[leftmargin=*]
	
	\item RNS~\cite{bpr} randomly samples negative items from the unobserved item set of each user.
	
	\item DNS~\cite{DNS} samples negative items from a candidate pool according to the current model-predicted preference scores.
	
	\item DNS(M,N)~\cite{DNSMN} is an extended version of DNS, which introduces predefined parameters to perform controllable hard negative sampling. The parameters $(M,N)$ are set to $(100,5)$ for Amazon-toys and Yelp, and $(50,10)$ for ML-1m.
	
	\item MixGCF~\cite{MixGCF} constructs a virtual hard negative pool by injecting positive-sample information into the candidate negative pool, and then selects the hardest negative according to the model-predicted matching scores.
	
	\item DENS~\cite{DENS} performs negative sampling by disentangling relevant and irrelevant factors, and selects negatives from the candidate pool that are close to positives in irrelevant factors but distinct in preference-relevant factors. 
	The weights for the relevant factor $\alpha$ and the gating task $\gamma$ are set to $(1.0, 0.3)$, $(0.0, 0.05)$, and $(0.0, 0.01)$ on Amazon-toys, Yelp, and ML-1m, respectively.

	\item BNS~\cite{BNS} first constructs a candidate pool and then selects negatives by jointly considering their informativeness and Bayesian-estimated likelihood of being true negatives, with the reliability weight $\lambda=5.0$ on all datasets.
	
	\item AHNS~\cite{AHNS} adaptively adjusts negative sampling hardness based on positive-sample information. The values of $\alpha$, $\beta$, and $p$ are set to 1.0, 0.1, and -2 on all datasets.
	
	\item DivNS~\cite{DivNS} samples diverse negatives from the cache via diversity-augmented $k$-DPP sampling. For Amazon-toys, Yelp, and ML-1m, the cache ratio $r$ and mixing coefficient $\lambda$ are set to $(4, 0.8)$, $(0, 1.0)$, and $(4, 0.8)$.
	
\end{itemize}

\subsubsection{Implementation Details}

Unless otherwise specified, we use LightGCN as the backbone. 
We initialize 64-dimensional embeddings with Xavier initialization~\cite{Xavier} and optimize the model using Adam~\cite{adam} with a learning rate of 0.001. The mini-batch size is 2048, the L2 regularization coefficient is 0.0001, and the number of layers is set to 3. Early stopping is performed based on validation Recall@20, with the patience value set to 10. All experiments are repeated with five random seeds, and we report the mean and standard deviation.
Our source code is available in https://anonymous.4open.science/r/SAHC-NS-CEF2/.

\subsection{Performance Comparison (RQ1)}

\begin{figure}[t]
	
	\includegraphics[width=\linewidth]{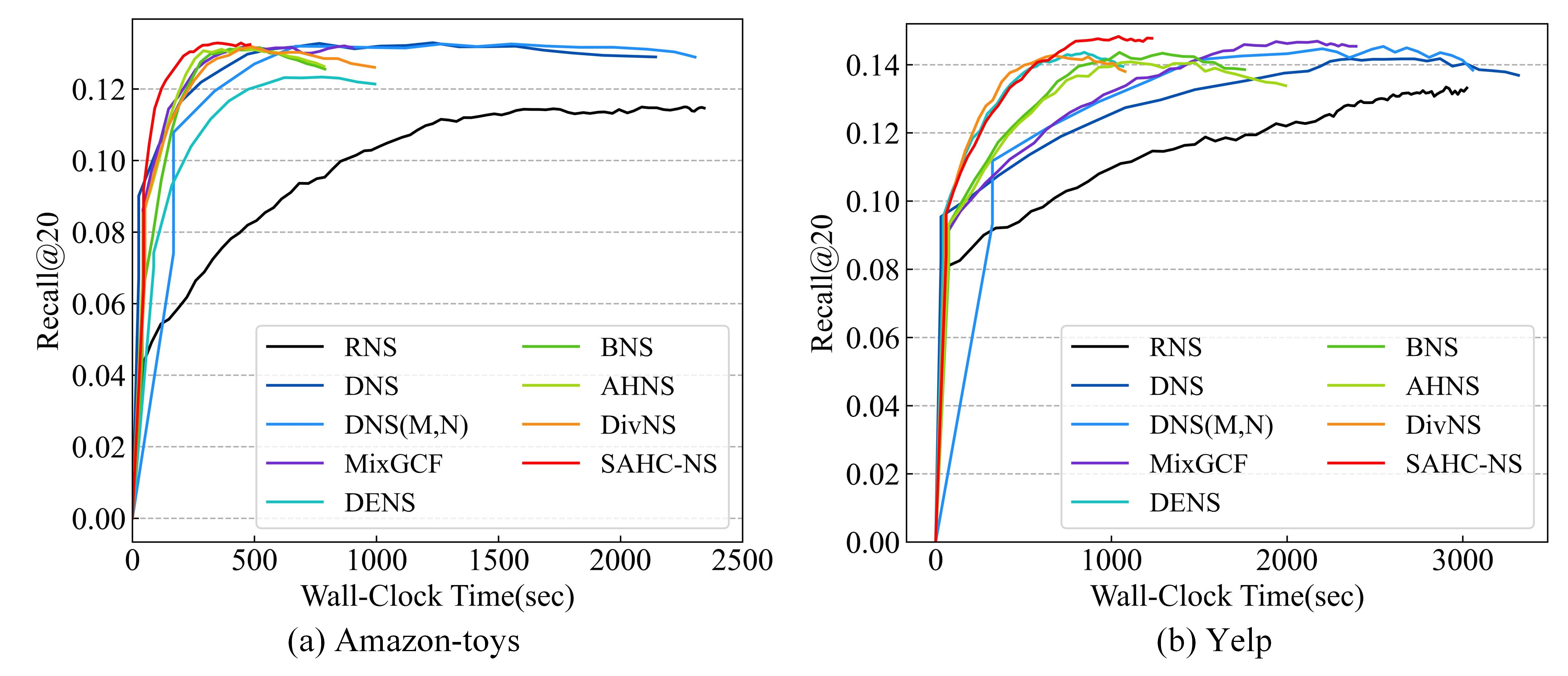}
	
	\vspace{-0.3cm}

	\caption{Recall@20 vs. wall-clock time (in seconds).}
	\label{shoulian}
	
	\vspace{-0.3cm}
	
\end{figure}

Table~\ref{tab:main} reports the performance of SAHC-NS and all baselines on three real-world datasets. Results are reported as mean $\pm$ standard deviation over five runs. We also conduct paired $t$-tests based on the results from identical random seeds across different methods to verify the statistical significance of performance improvements.
Overall, SAHC-NS achieves the best performance on all datasets and all evaluation metrics, demonstrating its effectiveness. 
This consistent improvement can be attributed to the complementary effects of the two modules in SAHC-NS. First, the SA module evaluates candidate negatives using layer-wise matching scores rather than relying only on final aggregated embeddings. By jointly considering mean hardness and cross-layer structural discrepancy, SAHC-NS selects negatives that are not only hard on average but also exhibit informative structural discrepancies across propagation layers, thereby capturing structural ambiguity that is overlooked by final-score-based sampling. 
Second, HC calibrates the augmentation strength of the negative selected by SA according to candidate-pool hardness, rather than relying on a fixed augmentation-strength distribution. 
For relatively easy pools, it increases the hardness and informativeness of the selected negative; for already hard pools, it avoids excessive hardening of the selected negative, thereby mitigating the potential false-negative risk associated with overly hard negatives.
As a result, SAHC-NS provides structure-aware and hardness-controllable negative samples, leading to more effective positive-negative discrimination during training.

\begin{figure*}[t]
	
	\includegraphics[width=\linewidth]{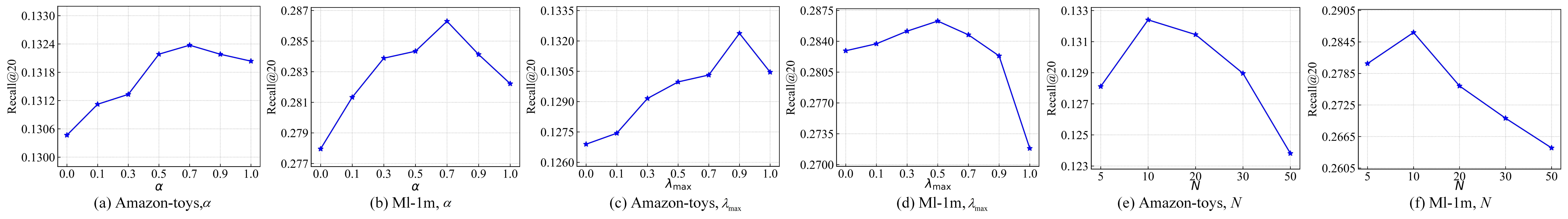}
	
	\vspace{-0.3cm}
	
	\caption{The impact of $\alpha$, $\lambda_{\max}$, and $n$. The y-axes are zoomed in to better illustrate performance variations.}
	
	\vspace{-0.3cm}
	
\end{figure*}

Beyond effectiveness, we further analyze the efficiency of SAHC-NS.
We first evaluate the overhead introduced by the SA module under the setting where HC is not used.
When SA is disabled, the sampler adopts standard hard negative sampling, i.e., selecting the negative item with the highest preference score from the candidate pool.
Enabling SA introduces layer-wise matching statistics for structure-aware negative selection, increasing the average per-epoch forward-pass time from 0.071s, 0.079s, and 0.137s to 0.136s, 0.138s, and 0.235s on Amazon-toys, Yelp, and ML-1m, respectively.
After further enabling HC, the corresponding time increases to 0.204s, 0.240s, and 0.400s.
Compared with the average overall training time per epoch, i.e., 6.26s, 11.89s, and 38.75s on the three datasets, the additional overhead introduced by SA and HC is small.
Specifically, the total extra sampler forward-pass time introduced by both modules is only 0.133s, 0.161s, and 0.263s, accounting for about 2.12\%, 1.35\%, and 0.68\% of the overall per-epoch training time.
This is because both modules operate only over a small candidate pool and mainly involve candidate-level inner products, simple score statistics, and interpolation-based calibration. These results indicate that the additional computational cost introduced by SAHC-NS is practically acceptable.
As shown in Fig.~3, SAHC-NS can achieve better performance within a relatively short training time.
Although SAHC-NS introduces additional computations, these operations are performed over the bounded candidate pool and are therefore lightweight.
Consequently, SAHC-NS achieves a favorable trade-off between recommendation performance and training efficiency.

\subsection{Parameter Analysis (RQ2)}

In this section, we investigate the sensitivity of SAHC-NS to three key parameters: $\alpha$, $\lambda_{\max}$, and the candidate-pool size $N$.
Here, $\alpha$ controls the contribution of cross-layer structural discrepancy in structure-aware negative selection, while $\lambda_{\max}$ controls the maximum calibration strength in HC.
The candidate-pool size $N$ determines the number of candidate negatives available for structure-aware selection and affects the hardness estimation of the candidate pool in HC.
We vary $\alpha$ and $\lambda_{\max}$ in $\{0.0, 0.1, 0.3, 0.5, 0.7, 0.9, 1.0\}$ and investigate the impact of candidate-pool sizes $N \in \{5, 10, 20, 30, 50\}$.

Fig.~4(a) and Fig.~4(b) show the impact of $\alpha$ on model performance.
As $\alpha$ increases, the performance first improves because a moderate $\alpha$ allows the SA module to incorporate cross-layer structural discrepancy in addition to mean hardness, helping the sampler identify negatives with informative layer-wise matching variations.
However, when $\alpha$ is too large, the selection score may place excessive emphasis on cross-layer discrepancy, causing the sampler to select candidates with large layer-wise fluctuations but insufficient average hardness, thereby weakening the training signals.

Fig.~4(c) and Fig.~4(d) show the impact of $\lambda_{\max}$ on model performance.
As $\lambda_{\max}$ increases, the performance first improves because a moderate $\lambda_{\max}$ enables HC to provide sufficient hardness enhancement while keeping the generated negatives hardness-controllable.
However, when $\lambda_{\max}$ is too large, the selected negative may be moved excessively toward the positive item, causing the generated negative to become overly hard and increasing potential false-negative risk.
As a result, the training signal becomes less reliable, leading to performance degradation.

Fig.~4(e) and Fig.~4(f) show the impact of the candidate-pool size $N$.
As $n$ increases, the performance first improves, since a larger candidate pool provides more candidate negatives and increases the likelihood of including informative samples for structure-aware selection and hardness calibration.
However, when $n$ becomes excessively large, the pool becomes more likely to contain extremely hard candidates, which may carry a higher false-negative risk.
Moreover, because HC estimates candidate-pool hardness using the highest-scoring candidate, an excessively large pool increases the chance that the hardness estimate is dominated by an extreme candidate.
This may cause HC to underestimate the amount of additional calibration required by the SA-selected negative, resulting in overly conservative hardness calibration and degraded performance.

\begin{table}[t] 
	\centering 
	\caption{Ablation study of different variants.} 
	
	\vspace{-0.2cm}
	
	\label{tab:ablation} 
	
	\begin{tabular}{ccccc} 
		\toprule 
		\multirow{2}{*}{Method}  
		& \multicolumn{2}{c}{Amazon-toys}  
		& \multicolumn{2}{c}{Yelp} \\  
		\cmidrule(lr){2-3} \cmidrule(lr){4-5} 
		& R@20 & N@20 & R@20 & N@20 \\  
		\midrule 
		
		DNS       & 12.66 & 7.14 & 13.67 & 7.15 \\ 
		SA-only   & 13.00 & 7.42 & 13.98 & 7.47 \\ 
		HC-only   & 13.04 & 7.53 & 14.24 & 7.56 \\ 
		
		\midrule 
		
		SAHC w/o Discrepancy & 13.01 & 7.46 & 14.42 & 7.57 \\ 
		SAHC w/o Mean        & 12.04 & 6.95 & 13.68 & 7.11 \\ 
		SAHC-Fixed           & 13.10 & 7.33 & 14.21 & 7.46 \\ 
		SAHC-Global          & 13.08 & 7.54 & 14.25 & 7.39 \\ 
		
		\midrule 
		\method            & \textbf{13.24} & \textbf{7.65} & \textbf{14.63} & \textbf{7.72} \\ 
		\bottomrule 
	\end{tabular} 
	
	\vspace{-0.5cm} 
	
\end{table}

\subsection{Ablation and Mechanism Study (RQ3)}

In this section, we conduct ablation experiments on the Amazon-toys and Yelp datasets. We design several variants at two levels: module-level ablation and mechanism-level ablation. The module-level ablation examines the necessity and complementarity of the SA and HC modules, while the mechanism-level ablation further investigates the effectiveness of the key designs within each module.
For module-level ablation, we compare SAHC-NS with DNS, SA-only, and HC-only. 
\textbf{DNS} removes both SA and HC and follows standard hard negative sampling by selecting the highest-scoring negative from the candidate pool; 
\textbf{SA-only} keeps SA but removes HC, directly using the negative selected by SA for training; 
\textbf{HC-only} removes SA and applies candidate-pool-aware hardness calibration to the negative selected by standard hard negative sampling. 
For mechanism-level ablation, we further construct SAHC w/o Discrepancy, SAHC w/o Mean, SAHC-Fixed, and SAHC-Global. 
\textbf{SAHC w/o Discrepancy} removes the cross-layer structural discrepancy term and selects negatives only according to mean hardness, whereas \textbf{SAHC w/o Mean} removes mean hardness and relies only on cross-layer structural discrepancy. 
\textbf{SAHC-Fixed} samples augmentation strengths from a fixed uniform distribution instead of adapting them according to candidate-pool hardness, \textbf{SAHC-Global} replaces layer-wise hardness estimation with a global hardness estimate computed from the final embeddings, and uses it to calibrate the augmentation strength across all propagation layers.

Table~\ref{tab:ablation} reports the ablation results of SAHC-NS.
Compared with DNS, SA-only consistently performs better, showing that structure-aware selection based on layer-wise matching scores is more effective than relying solely on the final matching score.
Specifically, by jointly considering mean hardness and cross-layer structural discrepancy, SA selects negatives that are both hard on average and structurally informative across propagation layers.
HC-only also outperforms DNS, showing that candidate-pool-aware hardness calibration is beneficial even for DNS-selected negatives.
These results confirm the effectiveness of both modules.
Furthermore, SAHC-NS outperforms both SA-only and HC-only, verifying the complementarity of the two modules. 
The improvement over SA-only shows that calibrating the hardness of SA-selected negatives is beneficial, while the gain over HC-only indicates that structure-aware selection provides better negatives before calibration. 
Thus, SA determines which negative should be selected, whereas HC controls how hard the selected negative should be made.
For the mechanism-level ablation, SAHC w/o Mean shows a larger performance drop than SAHC w/o Discrepancy, indicating that mean hardness is the dominant signal for selecting informative negatives.
However, SAHC w/o Discrepancy underperforms SAHC-NS, suggesting that cross-layer structural discrepancy helps identify structurally challenging negatives that cannot be fully captured by mean hardness alone.
SAHC-Fixed performs worse than SAHC-NS, showing that candidate-pool-aware calibration is more effective than using a fixed augmentation-strength distribution.
SAHC-Global also lags behind SAHC-NS, indicating that adaptively calibrating augmentation strength according to layer-wise candidate-pool hardness is more effective than using a global hardness estimate derived from the final embeddings.
Overall, these results verify the necessity and complementarity of the proposed SA and HC designs.

\begin{figure}[t]
	
	\includegraphics[width=\linewidth]{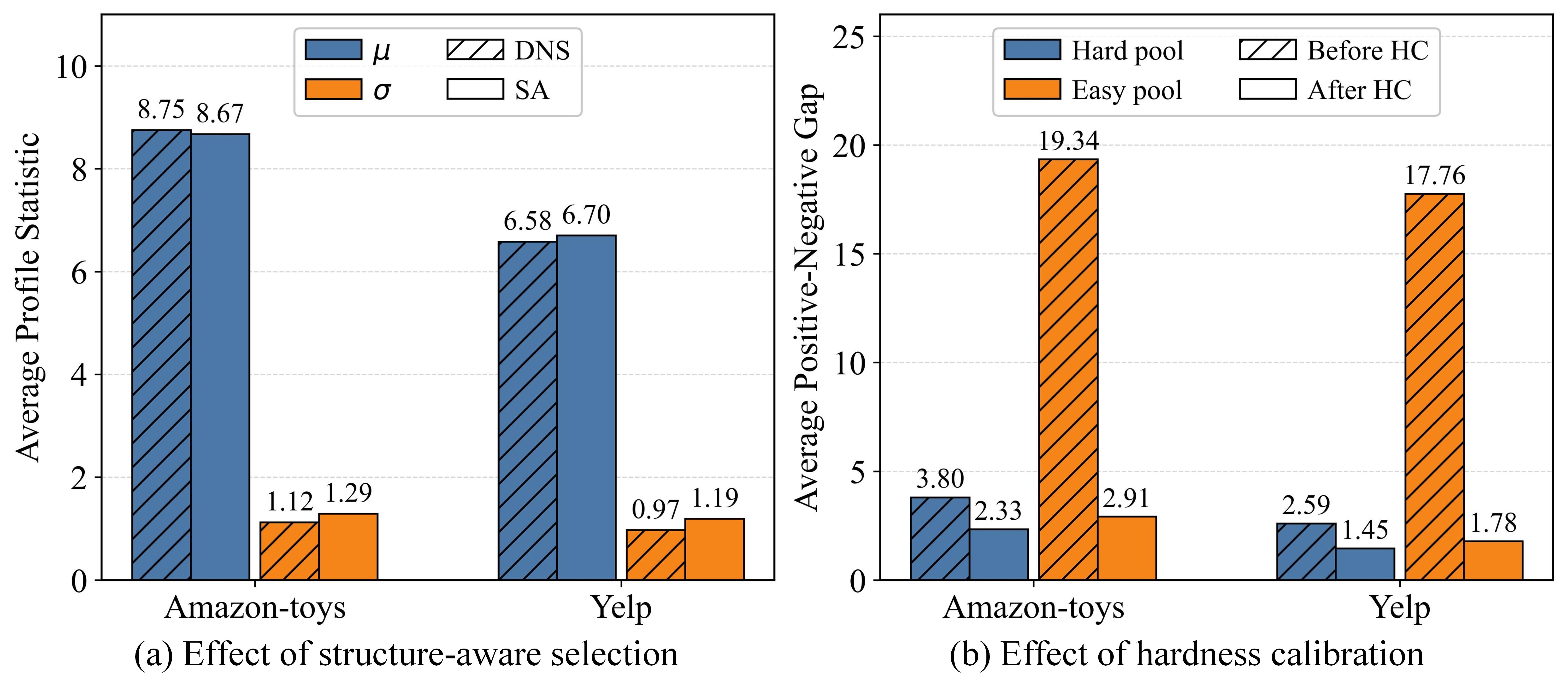}
	
	\vspace{-0.3cm}
	
	\caption{Mechanism analysis of the SA and HC modules.}
	
	\vspace{-0.5cm}
	
\end{figure}

Beyond the ablation study, we further conduct mechanism analyses to understand how SA and HC work in practice.
\textbf{Effect of structure-aware selection.}
On Amazon-toys and Yelp, for each training pair, we compare the negatives selected by DNS and SA from the same candidate pool and compute their mean hardness $\mu_{u,j}$ and cross-layer structural discrepancy $\sigma_{u,j}$.
Since the two statistics have different numerical scales, we focus on their relative changes.
As shown in Fig.~5(a), compared with DNS, SA selects negatives with larger cross-layer structural discrepancy while maintaining comparable mean hardness.
Specifically, the average discrepancy increases by 15.18\% and 22.68\% on Amazon-Toys and Yelp, respectively, while mean hardness varies by no more than 1.9\% on both datasets.
This indicates that SA does not simply select candidates with large layer-wise fluctuations, but identifies negatives that are hard on average while exhibiting more informative layer-wise matching variations.
These results show that SA effectively captures cross-layer structural discrepancies overlooked by DNS.
\textbf{Effect of hardness calibration.}
We then analyze how HC calibrates negative hardness under different candidate-pool conditions.
For each candidate pool, we compute the average positive-negative gap between the positive item and the hardest candidate negative across propagation layers, and treat the top 20\% and bottom 20\% of the averaged gap values as easy and hard pools, respectively.
Then, we compare the positive-negative gap of the selected negative before and after HC.
As shown in Fig.~5(b), calibrated negatives from easy pools exhibit a larger reduction in the positive-negative gap.
This suggests that HC applies stronger calibration to easy pools to enhance negative informativeness, while using milder adjustments for already hard pools to avoid excessive hardening and mitigate the associated potential false-negative risk.

\subsection{Applicability Analysis (RQ4)}

\begin{table}[t] 
	\centering 
	\caption{Performance comparison with various backbones.} 
	\label{tab:backbone} 
	\vspace{-0.1cm} 
	
	\setlength{\tabcolsep}{2pt} 
	\resizebox{\linewidth}{!}{ 
		\begin{tabular}{cccccc} 
			\toprule 
			\multirow{2}{*}{Backbone} & \multirow{2}{*}{Method} & \multicolumn{2}{c}{Amazon-toys} & \multicolumn{2}{c}{Yelp} \\ 
			\cmidrule(lr){3-4} \cmidrule(lr){5-6} 
			&& Recall@20 & NDCG@20 & Recall@20 & NDCG@20 \\ 
			\midrule 
			\multirow{9}{*}{NGCF} 
			& RNS      & 8.65 & 4.55 & 11.13 & 5.57 \\ 
			& DNS      & 11.24 & 6.00 & 13.39 & 6.99 \\ 
			& DNS(M,N) & 11.17 & 5.98 & 13.55 & 6.73 \\ 
			& MixGCF   & 11.61 & 6.22 & 13.52 & 7.03 \\ 
			& DENS     & 10.04 & 5.36 & 12.28 & 6.34 \\ 
			& BNS      & 10.61 & 5.58 & 12.80 & 6.50 \\ 
			& AHNS     & 10.34 & 5.30 & 12.66 & 6.61 \\ 
			& DivNS    & 10.90 & 5.43 & 12.07 & 5.97 \\ 
			& SAHC-NS  & \textbf{11.73} & \textbf{6.50} & \textbf{14.01} & \textbf{7.29} \\ 
			
			\midrule 
			
			\multirow{9}{*}{SimGCL} 
			& RNS      & 11.86 & 6.51 & 14.77 & 7.78 \\ 
			& DNS      & 12.95 & 7.41 & 14.52 & 7.66 \\ 
			& DNS(M,N) & 13.22 & 7.60 & 14.49 & 7.63 \\ 
			& MixGCF   & 13.32 & 7.60 & 15.02 & 7.92 \\ 
			& DENS     & 11.84 & 6.36 & 13.98 & 7.39 \\ 
			& BNS      & 12.12 & 6.87 & 13.87 & 7.28 \\ 
			& AHNS     & 12.71 & 7.20 & 14.67 & 7.76 \\ 
			& DivNS    & 11.98 & 6.77 & 13.98 & 7.34 \\ 
			& SAHC-NS  & \textbf{13.53} & \textbf{7.82} & \textbf{15.45} & \textbf{8.07} \\ 
			
			\bottomrule 
	\end{tabular}} 
	
	\vspace{-0.3cm} 
	
\end{table}

In this section, we apply SAHC-NS to other representative GNN-based implicit CF backbones, including NGCF~\cite{NGCF} and SimGCL~\cite{SimGCL}, to examine its generality.
Table~\ref{tab:backbone} reports the performance of different negative sampling methods under NGCF and SimGCL.
SAHC-NS achieves the best performance, demonstrating that it can be effectively generalized to different GNN-based recommendation backbones.
Moreover, stronger backbones generally improve the performance of most negative sampling methods, since higher-quality user and item representations provide more reliable preference signals for model-aware negative sampling.

\section{Conclusion}

In this paper, we studied negative sampling for implicit CF and identified two limitations of existing two-pass sampling methods: candidate-pool hardness mismatch and layer-wise structural blindness. To address these issues, we proposed SAHC-NS, a structure-aware and hardness-calibrated negative sampling method for GNN-based recommenders. SAHC-NS first performs structure-aware negative selection by modeling both the mean hardness and cross-layer structural discrepancy of candidate negatives through layer-wise matching scores. It then conducts candidate-pool-aware hardness calibration to adaptively adjust the augmentation strength according to the hardness of the candidate pool, thereby generating negatives that are informative while avoiding excessive hardness. Extensive experiments demonstrate that SAHC-NS achieves superior recommendation performance, maintains practical training efficiency, and generalizes well to different GNN-based recommenders.
These results suggest that jointly considering layer-wise structural discrepancy and candidate-pool hardness is a promising direction for improving negative sampling in implicit CF.

\bibliographystyle{IEEEtran}
\bibliography{sahcns_refs1}

\end{document}